# Brief Review of Recent Advances in Understanding Dark Matter and Dark Energy

Eugene Oks*

Physics Department, 380 Duncan Drive, Auburn University, Auburn, AL, USA

Abstract

Dark sector, constituting about 95% of the Universe, remains the subject of numerous studies. There are lots of models dealing with the cause of the effects assigned to "dark matter" and "dark energy". This brief review is devoted to the *very recent* theoretical advances in these areas: only to the advances achieved in the last few years. For example, in section devoted to particle dark matter we overview recent publications on sterile neutrinos, self-interacting dark matter, dibarions (hexaquarks), dark matter from primordial "bubbles", primordial black holes as dark matter, axions escaping from neutron stars, and dark and usual matter interacting via the fifth dimension. We also overview the second flavor of hydrogen atoms: their existence was proven by analyzing atomic experiments and is also evidenced by the latest astrophysical observations of the 21 cm spectral line from the early Universe. While discussing non-particle models of the cause of dark matter effects, we refer to modified Newtonian dynamics and modifications of the strong equivalence principles. We also consider exotic compact objects, primordial black holes, and retardation effects. Finally, we review recent studies on the cause of "dark energy effects". Specifically, we cover two disputes that arose in 2019 and 2020 on whether the observations of supernovas, previously interpreted as the proof of the existence of dark energy, could have alternative explanations. Besides, we note a study of 2021, where dark energy is substituted by a new hypothetical type of dark matter having a magnetic-type interaction. We also refer to the recent model of a system of nonrelativistic neutral gravitating particles providing an alternative explanation of the entire dynamics of the Universe expansion – without introducing dark energy or new gravitational degrees of freedom.

* Corresponding author: goks@physics.auburn.edu

## 1. Introduction

Dark sector, constituting about 95% of the Universe, remains the subject of numerous studies. There are lots of models dealing with the cause of the effects assigned to "dark matter" and "dark energy" – see, e.g., reviews by Salucci et al (2020), de Martino et al (2020), Frusciante & Perenon (2020), Tawfik & Dahab (2019), Bertone & Hooper (2018), Kenath et al (2017), and Joyce et al (2016), as well as references from these reviews.

Dark matter hypotheses can be subdivided in the following two groups. The first group is particle dark matter. In this group, some kind of a new particle (that vary from one hypothesis to another) is nominated as dark matter.

The second group consists of alternative hypotheses that try to claim that particle dark matter does not exist. First of all, these are modified gravity models. In these models, the authors propose various modifications of known physical laws while trying to explain the "dark matter effects" without inventing new particles. In this group there are also more exotic hypotheses.

As for the cause of "dark energy effects", the primary purpose of corresponding studies is to explain the observed accelerated expansion of the Universe at the present era. The most studied hypotheses seem to be the cosmological constant (the zero-point radiation of vacuum) and quintessence. More exotic hypotheses are, e.g., phantom energy, quintom, variations of physical constants, and modified gravity.

In subsequent sections we provide the corresponding references. Namely, we give recent references relevant to all of the above hypotheses on the cause of dark matter effects and dark energy effects.

The present brief review is devoted to the *very recent* theoretical advances in these areas: only to the advances achieved in the last few years. For example, in section devoted to particle dark matter we overview recent publications on sterile neutrinos, self-interacting dark matter, dibarions (hexaquarks), dark matter from primordial "bubbles", primordial black holes as dark matter, axions escaping from neutron stars, and dark and usual matter interacting via the fifth dimension. We also overview the second flavor of hydrogen atoms: their existence was proven by analyzing atomic experiments and is also evidenced by the latest astrophysical observations of the 21 cm spectral line from the early Universe. For these reasons, the second flavor of hydrogen atoms appears as the most likely explanation of dark matter – in distinction to the other hypotheses that either invented never-discovered particles or significantly changed the known physical laws.

While discussing the second group of models of the cause of dark matter effects, we refer to modified Newtonian dynamics and modifications of the strong equivalence principles. We also consider problems that modified gravity models face, as well as the recent study providing some observational evidence in favor of modified gravity. We also discuss exotic compact objects, primordial black holes, and retardation effects.

Finally, we review recent studies on the cause of "dark energy effects". Specifically, we cover two disputes that arose in 2019 and 2020 on whether the observations of supernovas, previously interpreted as the proof of the existence of dark energy, could have alternative explanations. Besides, we note a study of 2021, where dark energy is substituted by a new hypothetical type of dark matter having a magnetic-type interaction. We also refer to the recent model of a system of nonrelativistic neutral gravitating particles providing an alternative explanation of the entire dynamics of the Universe expansion – without introducing dark energy or new gravitational degrees of freedom.

## 2. Particle dark matter

Sterile neutrinos

Up to now, only left-handed neutrinos have been observed (their momentum and spin are antiparallel). Right-handed neutrinos interacting only gravitationally (in particular, not interacting electroweakly) is a hypothesis that goes beyond the Standard Model. Under this assumption, this kind of right-handed neutrino (called sterile neutrinos) is suggested as the dark matter particle.

All other known fermions have been found to be both right-handed and left-handed. So, it was kind of natural to propose a possible existence of right-handed neutrinos (Drewes, 2013).

Recently Boyle et al (2018) brought up the following additional considerations on the hypothetical right-handed neutrinos. In that paper, the authors proposed the existence of a "mirror Universe" that preceded the Big Bang. If our expanding Universe would be pictured as a cone, then the mirror Universe could be pictured as the symmetric continuation of this cone through its tip, which is the point of the Big Bang (see Fig. 1). While the second cone expands in the opposite direction of the time, the causality would not be violated within the mirror Universe.

In our Universe without its mirror counterpart, the Charge-Parity-Time (CPT) symmetry is broken. The double-cone design would restore the CPT symmetry.

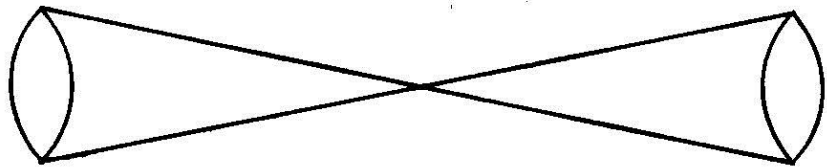

Fig. 1. Schematic of two "mirror Universes" discussed by Boyle et al (2018).

This design is related to the right-handed neutrinos hypothesis, as follows. Boyle at al (2018) theorized that in this kind of cosmology, the three known types of left-handed neutrinos, should have, as the counterparts, three types of right-handed neutrinos in our Universe. Further, according to Boyle et al (2018), only one (out of three) types of right-handed neutrinos would survive till now and (if sterile) represent dark matter particles. The authors estimated that in order to fit the observed density of dark matter, the surviving right-handed neutrinos should have the mass about $4.8 \times 10^8$ GeV = 480 PeV.

In the subsequent paper by Anchordoqui et al (2018), there was a suggestion that the ANITA team observations published by Gorham et al (2016), might be consistent with the existence of the right-handed neutrinos of the mass 480 PeV. The ANITA (Antarctic Impulsive Transient Antenna) system detected particles of the energies in the range of (600±400) PeV), presumably streaming out of the Antarctic.

This hypothesis faces the following unresolved problem: how these particles got close to Antarctica. The modelling suggests that these hypothetical right-handed neutrinos would fall to the Earth center after the collision with the Earth and thus should not fit the Anita observations.

Self-interacting dark matter

This hypothesis is the opposite of cold dark matter. The central assumption in this hypothesis, brought up by Spergel & Steinhardt (2000), is a strong interaction of dark matter particles with each other. Later such interaction was modeled by Yukawa-type potential (Loeb & Weiner (2005)).

Spergel & Steinhardt (2000) suggested that this interaction is facilitated by a light intermediate vector boson. Yet the mass of the vector boson must be large enough so that there should be no dissipation during the process of the scattering of dark matter particles.

The primary difference between the hypotheses of the self-interacting dark matter and cold dark matter is the following. According to the self-interacting dark matter hypothesis, dark matter particles experience some unknown dark force. In distinction, according to the cold dark matter model, dark matter particles do not engage in collisions.

Recently Yang et al (2020) performed two kinds of simulations of the dark matter distributions in two ultra-diffuse galaxies NGC 1052-DF2 and -DF4: one simulation based on the cold dark matter model, another based on the self-interacting dark matter hypothesis. Their simulations based on the cold dark matter model demonstrated that there is less tidal stripping of the inner halo of these two galaxies compared to the corresponding results based on the self-interacting dark matter hypothesis. The reason is that due to the self-interactions, the mass loss in the inner halo increases because dark matter particles are pushed from the inner halo to the outer parts of the galaxy. Since the recent observational data from the two ultra-diffuse galaxies NGC 1052-DF2 and -DF4 showed dark matter deficiency, Yang et al (2020) interpreted their results to be in favor of self-interacting dark matter. However, it should be emphasized that they had to make several assumptions in their simulations because the dark force acting on dark matter particles was (and is) unknown.

Another candidate for self-interacting dark matter is Bose-Einstein condensate – the condensate formed during the cosmological evolution in a bosonic gas (as first suggested by Membrado et al (1989)) – see, e.g., a recent paper by Craciun & Harko (2020) and references therein. Craciun & Harko (2020) studied the properties of the galactic rotation curves in the Bose-Einstein condensate dark matter model, with quadratic self-interaction – by using 173 galaxies from the Spitzer Photomery & Accurate Rotation Curves (SPARC) data. They found that this model gives a good description of the SPARC data. They also noted that regrettably the Bose–Einstein condensate dark matter model in its present formulation cannot provide any firm prediction on the mass of the dark matter particle.

Dibarions (hexaquarks)

Hypothetical dibaryons are also known as hexaquarks. This is because hypothetically dibaryons should have six quarks or antiquarks regardless of the flavor.

Adlarson et al (2014) claimed that, while performing experiments on the quasifree polarized $np$ scattering, they observed a narrow resonance at 2380 MeV. The theoretical prediction of this kind of resonance was due to Dyson & Huang (1964). Goldman et al (1989) gave it the name d*. This hypothetical particle d*(2380) is presumed to have three up quarks and three down quarks.

The claim by Adlarson et al (2014) was contested by Bugg (2014). He indicated that Adlarson et al (2014) interpretation of the experiments has logical flaws. Bugg (2014) showed that the observation by Adlarson et al (2014) could be explained as the reaction $pd \rightarrow NN^*$

(1440)*ps*, where *ps* is a spectator proton, and also pointed out that the experiment by Mielke et al (2014) on $dp \to {}^3He\ \pi+\pi$ supports his interpretation.

Recently Bashkanov & Watts (2020) suggested that the system, consisting of hexaquarks with trapped electrons, could evolve in Bose-Einstein condensate. They theorized that in the early Universe, characterized by low temperatures, hexaquarks could overlap, attach to each other and, in such a state, would represent dark matter. By using a liquid drop model, they calculated some properties of such exotic Bose-Einstein condensate.

Siegel (2020b) criticized Baskanov-Watts' hexaquark-made Bose-Einstein condensate. His main argument was that this exotic condensate (even if created) would not survive very intense radiation that occurred in the early Universe. This very intense radiation would break apart the hexaquarks, and then there would be no way for hexaquarks to reassemble in the Bose-Einstein condensate: the necessary conditions for reassembling would not be there anymore.

Dark matter from primordial "bubbles"

Recently Baker et al (2020) hypothesized that dark matter could originate from a phase transition that occurred in the very early Universe. They suggested that, in addition to the known particles of the matter, in the very early Universe existed also unspecified dark matter particles of a variety of masses. Baker et al (2020) theorized that at that stage there occurred a formation of cooled plasma bubbles. In the course of the evolution, these bubbles presumably expanded and merged, thus constituting a phase transition.

During the course of the expansion of the bubbles, they "strained" dark matter particles into the bubbles (from the plasma). The walls of the bubbles performed a filtration: the lower mass dark matter particles were left outside and annihilated, but the higher mass dark matter particles got inside the bubbles and survived.

According to Baker et al (2020), this scenario could explain the amount of dark matter deduced from the current observations. However, in their model, dark matter particles should have massed from TeV to PeV: these would be very much heavier dark matter particles compared to other hypotheses.

Primordial black holes as dark matter

This hypothetical type of black holes could presumably form soon after the Big Bang – through the gravitational collapse in the heterogeneous high-density structure of the early Universe. This is the distinction of primordial black holes from the "usual" black holes, the latter being created by the stellar gravitational collapse. There were suggestions that primordial black holes could be dark matter. However, it turned out that for matching the known amount of dark matter it is problematic to develop the corresponding models yielding the necessary amount of primordial black holes.

Recently Carr & Kühnel (2020) considered a possibility that primordial black holes of masses greater than $10^3$ solar masses could resolve some cosmological problems even if they would constitute only a small fraction of the density of dark matter. Such primordial black holes could generate cosmological structures and remove some problems with the standard cold dark matter scenario.

Also Takhistov et al (2021) suggested that solar-mass black holes, which are not expected from conventional stellar evolution, can be produced by neutron star implosions induced by the capture of small primordial black holes or from accumulation of some varieties of particle dark matter, but the authors did not discuss what dark matter consists of. They theorized that data from future observations of gravitational waves would enable astrophysicists to distinguish between solar-mass black holes and neutron stars.

Axions escaping from neutron stars

The term "axion" is due to Wilczek (1978). These neutral spinless particles were first hypothesized by Peccei and Quinn (1977). These particles should have zero or little interaction with the ordinary matter and are supposed to have a very small mass – see, e.g., review by Luzio et al (2020) and references therein. We mention also recent papers from Balakin's group, where they studied, e.g., axionic extension of the Einstein-aether theory (Balakin & Shakirzyanov (2019)), features of axionically active plasma in a dyon magnetosphere (Balakin & Groshev (2019)), profiles of the axionic dark matter distribution (Balakin & Groshev (2020)), and whether the axionic dark matter is an equilibrium system (Balakin & Shakirzyanov (2020)), the model in the latter paper predicting an infinite number of equilibrium states.

Axions have never been found experimentally despite numerous attempts. Buschmann et al (2021) tried to connect axions to some astrophysical observations. Namely, Buschmann et al (2021) focused at the cores of neutron stars, where hypothetically, in the nucleon-nucleon scattering, axion bremsstrahlung takes place. From the core of neutron stars, these axions get out in the strong magnetic field of these stars, where there occurs their conversion in x-ray photons.

Dessert et al (2020) observed an excessive x-ray radiation in the 2-8 keV range from seven neutron stars and Buschmann et al (2021) tried to fit their model for explaining these oibservation. For this purpose, first of all, the axion mass in the model by Buschmann et al (2021) should be ~ $10^{-5}$ eV (or less). Second, there was a restriction on the axion-photon and axion-neutron coupling constants: their product should be between $2 \times 10^{-21}$/GeV and $10^{-18}$/GeV.

However, observations by Dessert et al (2020) could have alternative explanations, as noted by Buschmann et al (2021). One of the alternative explanations could related to an unknown process in neutron stars – the process different from the one proposed by Buschmann et al (2021). Another explanation could be that what was observed by Dessert et al (2020) was an artifact caused by the x-ray telescopes.

Moreover, according to the study by Ng et al (2021), it seems that dark matter actually cannot be represented by axions (or by light supersymmetry particles). The starting point of that study was that dark matter bosons (like axions), if exist, should be also present in the vicinity of black holes. For rotating black holes, capturing these particles would slow down the rotation. Ng et al (2021) analyzed the LIGO and Virgo data of black hole mergers, from which it is possible to deduce the rotation frequency of black holes before the merger. They found that some of these black holes had such a high rotation frequency that it was inconsistent with the hypothetical existence of axions or light supersymmetry particles with masses in the range between $1.3 \times 10^{-13}$ eV to $2.7 \times 10^{-13}$ eV. (The authors noted though that there are astrophysical scenarios that may explain the observed data without ruling out the existence of a boson in that mass range.)

Dark and usual matter interacting via the fifth dimension

Theories engaging the fifth dimension originated from papers by Kaluza (1926) and Klein (1926a, 1926b). About 70 years later there was the second wave of the interest to this concept. For example, according to the scenario by Randall & Sundrum (1999a, 1999b), all particles of the ordinary matter, except graviton, are confined to (3 + 1)-dimensional brane. As for graviton, it can move into the fifth dimension.

The thrust of the Randall-Sundrum's scenario was to try explaining why the gravitational force is b an astronomical number of times weaker than the weak force. In their model, there is a brane where the gravity is strong (Gravitybrane) and a brane, where all other particles of the ordinary matter are confined (Weakbrane). On the way from the Gravitybrane to the Weakbrane, the graviton's probability distribution falls off exponentially and thus becomes becomes very significantly weakened within the Weakbrane.

Carmona et al (2021), operating within the Randall-Sundrum scenario, suggested a new scalar field to exist: the field corresponding to a hypothetical heavy boson – in the analogy to the relation between the Higgs field and the Higgs boson. They theorized that the wave functions of these two kinds of bosons are entangled along the fifth dimension, so that the Higgs boson and this hypothetical boson could interact.

Carmona et al (2021) proposed dark matter particles to be 5-dimensional fermions, whose interaction with the ordinary matter is facilitated by this hypothetical boson. However, existing particle accelerators would not be able to the new heavy boson because, according to the estimates, its mass should be ~ 30 Tev.

These ideas by Carmona et al (2021) are interesting. However, their model requires a warped fifth dimension and two new types of particles operating in the fifth dimension, so that their model does not look favorable from the viewpoint of the Occam razor principle.

The second flavor of hydrogen atoms

Bowman et al (2018) published a perplexing observation of the redshifted 21 cm spectral line from the early Universe. The amplitude of the absorption profile of the 21 cm line, calculated by the standard cosmology, was by a factor of two smaller than it was actually observed. The consequence of thus striking discrepancy was that the gas temperature of the hydrogen clouds was actually significantly smaller than predicted by the standard cosmology.

The first attempt to explain this perplexing observation was undertaken by Barkana (2018). His hypothesis involved some unspecified dark matter colliding with the hydrogen gas and making it cooler compared to the standard cosmology. For fitting the observations by Bowman et al (2018), the mass of these dark matter particles should not exceed 4.3 GeV, as estimated by Barkana (2018).

Thereafter McGaugh (2018) examined the results by Bowman et al (2018) and Barkana (2018) and came to an important conclusion. Namely, the observations by Bowman et al (2018) constitute an *unambiguous proof that dark matter is baryonic*, so that models introducing non-baryonic nature of dark matter have to be rejected.

What if dark matter baryons, which provided an additional cooling to the primordial hydrogen gas, whose existence has been already proven – rather than being some yet unknown, unspecified particles, suggested by Barkana (2018)? This question has been posed and positively answered by Oks (2020a), as follows.

For dozens and dozens of years, there remained a mystery of a giant discrepancy between the theoretical and experimental results on the High-energy Tail of the linear Momentum Distribution (hereafter HTMD) in the ground state of hydrogen atoms.

Figure 2 shows the ratio of the theoretical HTMD, calculated by to Fock (1935), to the actual HTMD deduced by Gryzinski (1965) from the analysis of atomic experiments. It is seen that the relative discrepancy between the theoretical and experimental HTMD can reach of many orders of magnitude.

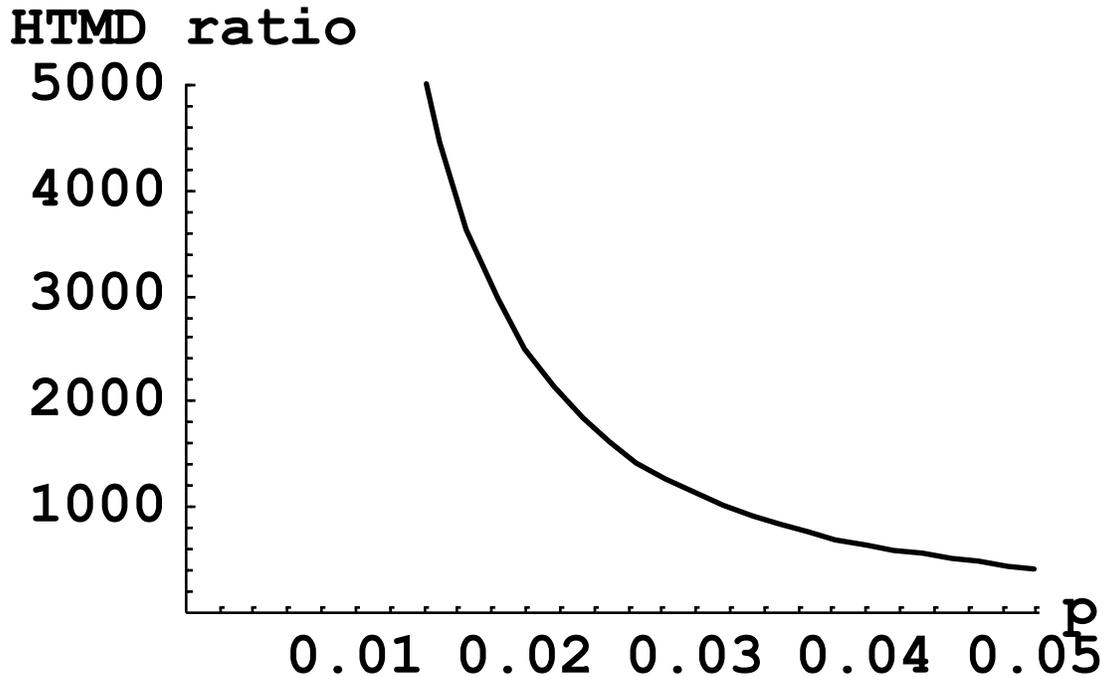

Fig. 2. Ratio of the theoretical High-energy Tail of the linear Momentum Distribution (HTMD) in the ground state of hydrogen atoms, calculated by Fock (1935), to the actual HTMD deduced by Gryzinski (1965) from the analysis of atomic experiments. The linear momentum p is in units of mc, where c is the speed of light.

The key to resolving the above giant discrepancy turned out to be in re-analyzing solutions of the Dirac equation for hydrogen atoms (and more generally for hydrogenic atoms/ions of a nuclear charge Z), which has two solutions. One solution has only a weak singularity as the radius-vector of the bound electron r → 0 and is called regular, while the other solution has a stronger singularity at r → 0 and is usually rejected.

Oks (2001) showed that with the allowance for a finite size of the nucleus of the hydrogen atom (the proton), for the ground state of hydrogen atoms, the *singular* solution of the Dirac equation outside the proton can be matched at the boundary with the regular solution of the Dirac equation inside the proton. Therefore, for the ground state of hydrogen atoms, the radial part of the Dirac bispinor should be a linear combination of the corresponding regular and *singular* solutions.

Oks (2001) demonstrated that this fundamental finding eliminated the above giant discrepancy between the theoretical and experimental HTMD in the ground state of hydrogen atoms. Thus, this constituted the experimental proof of the existence of an alternative kind of hydrogen atoms described by the singular solution of the Dirac equation outside the nucleus.

In paper by Oks (2020a) it was shown that the singular solution of the Dirac equation outside the proton is actually legitimate not only for the ground state of hydrogen atoms, but for all the states $n^2S_{1/2}$, where n is the principal quantum number ( n = 1, 2, 3, …). It is also legitimate for the $l = 0$ states of the continuous spectrum, where $l$ is the quantum number of the angular momentum.

In other words, there exists an alternative kind of hydrogen atoms – later called the Second Flavor of Hydrogen Atoms (SFHA) in Oks (2020b) paper – the hydrogen atoms having only discrete and continuum states of $l = 0$ and described by the singular solution of the Dirac equation outside the proton. Due to the quantum selection rules, the SFHA does not have excited discrete states that can be coupled to the ground state via the electromagnetic radiation – in distinction to usual hydrogen atoms.

The SFHA still possesses two hyperfine sublevels of the ground state and thus can engage in the formation of the 21 cm line signal. However, due to the absence of excited states that could be coupled to the ground state by the electric dipole radiation, the decoupling of the SFHA from the cosmic microwave background occurs earlier (in the course of the Universe expansion) than for the ordinary hydrogen atoms (Oks (2020a)). This is why the primordial hydrogen gas becomes significantly cooler than predicted by the standard cosmology. As a result, the giant discrepancy between the observations by Bowman et al (2018) and the standard cosmology gets eliminated – qualitatively and quantitatively.

The latest puzzling astrophysical observation is the following. The largest and most detailed map of the distribution of dark matter in the Universe has been recently created by the DES team (Chang et al 2018, Jeffrey et al 2021). The distribution turned out to be smoother than predicted by Einstein general relativity (by Einstein's gravity), according to Jeffrey et al (2021). As a result, there were calls for new physical laws: for some kind of non-Einsteinian gravitation. However, Oks (2021a) showed that this puzzling observation can be also explained both qualitatively and quantitatively by using the SFHA – without changing physical laws.

Namely, Oks (2021a) analyzed a system of a large number of gravitating neutral particles, whose mass is equal to the mass of hydrogen atoms, and focused at the subsystem of relatively isolated pairs of these particles. The pairs loose the energy by the gravitational radiation and the separation within the pair decreases. He showed that this process would stop as the separation of the particles within the pair would decrease to the minimum value of the order of few megaparsecs. This minimum value is practically the same as the average observed separation between galaxies.

The termination of the gravitational radiation and of the further decrease of the separation of the particles within the pairs is equivalent to a partial inhibition of the classical gravitation. He estimated that the percentage of the pairs, exhibiting the inhibition of the gravitational interaction and thus the inhibition of the unlimited "clumping", is $\gtrsim 2.5\%$. This agrees with the percentage observed by the DES team: the few percent more smooth, less clumpy distribution of dark matter compared to the prediction of the general relativity.

Dark matter particles having the mass of hydrogen atoms, interacting only gravitationally (but not electromagnetically) could only be SFHA. Thus, this constitutes the 2nd astrophysical evidence of the existence of the SFHA - in addition to the evidence of its existence from atomic experiments. This solidifies the status of the SFHA as the most probable candidate for dark matter or at least for a part of it.

In summary: the SFHA, as the candidate for the particle dark matter (by virtue of explaining the perplexing observations by Bowman et al (2018)), as well as the puzzling observations by Jeffrey et al (2021), has the following advantages compared to other candidates.

1. The SFHA has the *experimental proof* of the existence while other candidates do not.
2. All other candidates for the dark matter – not only for the particle dark matter, but in both categories of hypotheses – are beyond the Standard Model. In distinction, the SFHA is *within the Standard Model* and it is *based on the standard quantum mechanics*: the known physical laws have not been changed. For this reason, the principle of the Occam razor should be in favor of the SFHA.

At the end of this section, we note the following. As McGaugh (2018) demonstrated that the observations by Bowman et al (2018) clearly prove that dark matter is baryonic, Tatum (2020) hypothesized "cold atomic hydrogen in its ground state" might represent dark matter required for explaining the observations by Bowman et al (2018). However, Tatum (2020) disregarded the following. Ordinary hydrogen atoms, to which he referred, interact with the electric dipole radiation and thus can be excited to higher states (from the ground state) by absorbing a quantum of the electric dipole radiation. For this reason, the ordinary hydrogen atoms decouple from the cosmic microwave background at the stage predicted by the standard cosmology and therefore cannot justify the cooling to the temperature required for explaining the observations by Bowman et al (2018).

### 3. Non-particle dark matter

Modified gravity

Modified gravity is also known as "modified Newtonian dynamics". The motivation behind this hypothesis was to explain the observed "rotation curves", i.e., why the stars in most galaxies are observed to move at similar speeds regardless of their distance from the center of the galaxy. These observations are contrary to the prediction by Newtonian dynamics, according to which stars at the edge of a galaxy should move with smaller velocities compared to stars that are closer to the galactic center. For removing the discrepancy between the theory and the observed rotation curves, Milgrom (1983) proposed that at the edge of galaxies the gravitational force is weaker than in Newton gravity.

In later versions the modifications concerned the inertial mass. In the modified Newtonian dynamics, it was proposed that the inertial mass is the emergent (rather than inherent, as by Newton and Einstein) property of the object. Presumably, the net gravitational pull from the rest of the Universe (i.e., the external gravitational force) affects the inertial mass.

For the very latest developments, tests, and restrictions on various modified gravity hypotheses we refer to the following two reviews published in 2021 and references therein. The first review is by Baker et al (2021). It discusses, in particular, which modify gravity models survive the (so far) available astrophysical tests. Another review is by Chen et al (2021). They discuss modified gravity in the context of quasinormal modes of perturbed black holes.

The primary problem of modified gravity hypotheses is caused by the fact that galaxy clusters, even after being processed by means of the modified Newtonian dynamics, still show a residual mass discrepancy (McGaugh (2015)). So, these hypotheses cannot completely eliminate the need for dark matter to exist.

To conclude this brief discussion of the modified gravity we note the paper by Chae et al (2020) where they analyzed 175 galaxies as follows. If the modified gravity corresponds to reality, then the external field, caused by the distribution of nearby galaxies, should change the rotation curve of the galaxy under consideration. The authors found that such change was confirmed for galaxies, experiencing the strongest external field. However, there was no change in the rotation curve for galaxies experiencing a weaker external field. Thus, Chae et al (2020) did not disprove dark matter: a further study would be in order.

Exotic compact objects

Exotic compact objects is a term related to a group of proposals dealing with different hypothetical compact stars. They do not have electrons, neutrons, and protons. They are not collapsing by gravity presumably because of either degeneracy pressure or quantum properties – see, e.g., review by Cardoso & Pani (2019) and references therein.

One of these hypothetical objects has to do with *wormholes*. They were first considered by Einstein & Rosen (1935). Relatively recently it was understood that the stability of the wormholes can be achieved with some reasonable matter and could be traversable – see, e.g., paper by Anabalon et al (2020) and references therein.

Another hypothetical compact object is *quark star*. According to this hypothesis, first brought up by Ivanenko & Kurgdelaidze (1965), in the conditions of extreme pressure and temperatures, neutrons could dissolve into quarks, the latter becoming densely packed. The gravitational collapse of such quark stars would be prevented by the electric repulsion and the degeneracy pressure. There is no proof of their actual existence because extreme conditions required for their formation presently cannot be reproduced experimentally and have not been observed.

The next hypothetical compact object is *superspinar*. These are hypothetical objects introduced in frames of the string theory. Presumably they rotate so fast that their angular momentum is greater than it is admissible in the Kerr geometry – see. e.g., the paper by Gimon & Horava (2009).

Further, there was proposed an object called *fuzzball*. These objects defy the distinct, well defined event horizon of black holes. At the scale of comparable to the order of Plank length, the event horizon of these hypothetical objects would be fuzzy, thus justifying their name – see, e.g., paper by Bena & Warner (2013) and references therein. In this concept, the black hole geometry represents a superposition of microstates, whose event horizons slightly differ from one another, thus resulting in a fuzzy event horizon for the entire system.

The next hypothetical object is *gravastar*. This term is the abbreviated version of "gravitational-vacuum star" proposed by Mazur & Mottola in 2001 – see, e.g., their later paper Mazur & Mottola (2015) and references therein. Gravastars are supposed to have two distinct regions. The inner region of gravastars is the gravitational analog of Bose-Einstein condensate. In the outer region, i.e., outside the event horizon, they are supposed to look like the usual black holes – to have the same manifestations as the usual black holes.

Hypothetically, there could be also other horizonless configurations called "*2-2 holes*". Other hypotheses invoke *boson stars* (as opposed to the usual stars consisting mostly of fermions), *preon stars* (composed of hypothetical particles preons and having much higher density than even hypothetical quark stars), *Q stars* (extremely compact heavy neutron stars, characterized by an exotic equation of state and by the radii less than 3/2 of the Schwarzshield radius), *electroweak stars* (whose gravitational collapse is counteracted by the energy output of the process of the conversion of quarks into leptons, the process facilitated by the electroweak interaction), and so on – see, e.g., review by Cardoso & Pani (2019) and references therein.

Primordial black holes

The existence of these hypothetical type of black holes in the early Universe was suggested for the first time by Zel'dovich & Novikov (1966). The difference from the usual black holes is that the latter resulted from the gravitational collapse, while the hypothetical primordial black holes did not.

The latest twist of this story is the hypothesis that they resulted from vacuum bubbles through the process of quantum tunneling (Deng (2020), Deng & Vilenkin (2017)). According to this idea, "baby Universes" developed and split from the our Universe leading to the primordial black holes.

There have been suggested also "stupendously large black holes" (SLABs): the primordial black holes of much larger masses than any black hole observed so far. According to the paper by Carr et al (2021), if SLABs (themselves being not dark matter candidates) would be ever observed, this would increase the probability that light primordial black holes could exist as well. In this case, light primordial black holes might be dark matter candidates.

Retardation effects

Yahalom (2020) engaged retardation effects within general relativity for explaining the observed rotation curves of galaxies – without resorting to dark matter. This is an interesting idea from the theoretical point of view. However, it leaves without any explanation the gravitational lensing effect. So, the latter would still require introducing dark matter.

## 4. Dark energy

The primary purpose of various hypotheses about dark energy is to explain the observed accelerated expansion of the Universe at the present era. The simplest (and minimalistic) hypothesis is a constant energy density homogeneously distributed in the vacuum. This is known as the cosmological constant or, equivalently, the zero-point radiation of space – see, e.g., reviews by Kenath et al (2017) and Joyce et al (2016), as well as references from these reviews. While this hypothesis is consistent with all current observations, it has its intrinsic theoretical problem: the value of the cosmological constant that follows from observations is by 120 orders of magnitude smaller than the theoretically predicted value.

The next most popular hypothesis is the fifth fundamental force called quintessence, proposed by Peebles & Ratra (1988). It is a scalar field whose energy density undergoes a temporal and spatial evolution. The field should have a relatively large Compton wavelength (that is, to be relatively light) – to avoid forming structures/matter. Within the Standard Model, scalar fields must have relatively large masses. So, viable quintessence models should go beyond the Standard Model.

There are also more exotic hypotheses. One of them is phantom energy. It is characterized by a negative kinetic energy, such that as the Universe expands, the kinetic energy increases. Next, there is a hypothesis called quintom that is a hybrid of the quintessence and phantom hypotheses. In addition, there is also a hypothesis by Gurzadyan & Xue (2003), according to which physical constants (the gravitational constant, the speed of light) can vary. Besides, there are k-essence models (introduced by Armendariz-Picon et al (2001)), where the negative pressure (required for explaining the present era of the accelerated expansion of the

Universe) results from the nonlinear kinetic energy of the scalar field. According to these models, the negative pressure id exhibited only after the matter dominated the Universe for some time. One of the latest twists of the k-essence models was offered by Bandyopadhya & Chatterjee (2021). They studied the situation where there would be time-dependent diffusive interaction between dark matter and dark energy. They demonstrated that their model could be consistent with the observations of luminosity distance redshift data in Supernova Ia (SN Ia) observations.

Niedermann & Sloth (2021) proposed that in the early Universe there existed "extra dark energy" that presumably was a different phase of dark energy compared to its current state. The extra component of dark energy in the early Universe cause a short boost of additional repulsion. A phase transition occurred shortly before recombination era. This transition would decrease an initially high value of the cosmological constant in the early Universe down to the current value. This hypothesis could explain the discrepancy between the early and late measurements of the Hubble parameter.

Finally, there are hypotheses that dark energy does not exist. These hypotheses (such as modified gravity etc) provide alternative explanations to the currently observed accelerated expansion of the Universe – see, e.g., the review by Clifton et al (2012), the paper by Elmardi & Abebe (2017), as well as references from these publications.

One of the hypotheses called massive gravity is that gravitons have non-zero mass and/or leaks into a fifth dimension (resulting in the situation where at large distances gravity decreases faster than in the general relativity, thus causing the accelerated expansion of the Universe). One of the consequence would be for the gravitational waves propagate slower than the speed of light. However, observations did not confirm this. Besides, the massive gravity would lead to the graviton mass varying over time proportionally to the Universe expansion rate, which is weird from the theoretical point of view (Siegel, 2020a).

Other modified gravity hypotheses introduce various functions f(R) of the Ricci scalar – see, e.g., the review by Sotiriou & Faraoni (2010), the paper by Elmardi & Abebe (2017), as well as references from these publications. By introducing an arbitrary function, it could be possible at least in principle to explain the accelerated expansion of the Universe without resorting to dark energy. We note also recent papers by Odintsov's group where they studied, e.g., some aspects of the so-called f(R) gravity (Odintsov & Oikonomou (2020a), (2020b)). However, these hypotheses encounter their own stumbling blocks while trying to explain all observations.

Here is the bottom line. According to the review by Weinberg & White (2019), "*at present there are no fully realized and empirically viable modified gravity theories that explain the observed level of cosmic acceleration*".

The focus of the present review in general and of this section in particular is at *very recent* advances on the subject. With respect to dark energy, we bring the following recent advances to the attention of the research community.

In the paper by Colin et al (2019a) from Sarkar's group, the authors put in doubt the observational evidence of the accelerated expansion of the Universe. They argued that after they took into account the local motion of our telescopes, while analyzing the correlation between the red-shift and brightness of supernovas, the "monopole" (i.e., dark-energy-caused isotropic) accelerated expansion has a statistical significance of only $1.4\sigma$, while the "dipole-type" situation (presumably caused by the local motion) has a statistical significance of $3.9\sigma$.

In response, Rubin & Heitlauf (2019) pointed out four problems with the analysis by Colin et al (2019a). Then Colin et al (2019b) rebutted the Rubin-Heitlauf criticism stating that even after implementing corrections suggested by Rubin & Heitlauf, the evidence for isotropic acceleration rises to just 2.8σ and thus does not constitutes a sufficient proof.

Another dispute on the observational evidence of the existence of dark energy deduced by analyzing supernovas has been initiated by Kang et al (2020). While cross-analyzing the luminosity of supernovas and the ages of the stars in the corresponding galaxies, they found a significant correlation. This means that the luminosity of supernovas varies with the cosmic time: this is contrary to the opposite assumption made while discovering the existence of dark energy by analyzing supernovas.

The research community remained skeptical because the study by Kang et al (2020) demonstrated only a weak dependence of the supernovas luminosity with the age. Also, Kang et al (2020) took into account a smaller sample of galaxies compared to other studies, while including some supernovas that are older than the age of the Universe.

It should be emphasized that the above disputes concerns only the evidence of the existence of dark energy deduced from the analysis of supernovas. However, there is other observational evidence of dark energy effects, as deduced from the cosmic microwave background, the clustering of matter in the Universe, the gravitationally distorted shapes of galaxies, and baryon acoustic oscillations.

We note also an alternative suggestion by Loeve et al (2021). In their model, dark energy is replaced by a hypothetical new type dark matter exerting a velocity-dependent force similar to a magnetic force. This force was considered to be inversely proportional to the square of the distance and to be proportional to the velocity dispersion inside a galaxy. Loeve et al (2021) showed that under this hypothesis, i.e., under this *new physics*, they can explain the temporal evolution of the Universe without the need for the cosmological constant.

To conclude this section, we mention the paper by Oks (2021b) proposing an alternative explanation for the era of the decelerating expansion of the Universe and the era of the accelerated expansion of the Universe – without resorting to dark energy or without introducing new gravitational degrees of freedom. For the era of the decelerating expansion of the Universe, one of the central points was the application of Dirac's Generalized Hamiltonian Dynamics (Dirac (1950), (1958), (1964), Oks & Uzer (2002) to a system of nonrelativistic neutral gravitating particles, while another central point was the application of the generalized virial theorem (Polard (1964)).

Then there was pointed out the paper by Shamir (2020) where the author draw attention to the observed disparity between galaxies rotating counterclockwise and clockwise. From this observation followed the rotation of the Universe as the whole. Oks (2021b) underlined that while taking into account the results of the paper by Shamir (2020), the particles in the considered system would undergo an *expansion at an increasing speed* due to the centrifugal force.

For about the first 9 billion years, the Universe expansion was decelerating (before starting to accelerate about 5 billion years ago), as it is known (see. Fig. 3). During the first 9 (out of 12) stages of Oks (2021b) model, the gravitation prevailed over the centrifugal force, thus causing the Universe expansion to slow down. Then at stage 10, the gravitation got partially inhibited and the centrifugal force began to prevail, so that the expansion of the Universe started to accelerate. In this way, the entire history of the Universe expansion – both the deceleration and acceleration epochs – have been explained without resorting to dark energy or without introducing new gravitational degrees of freedom.

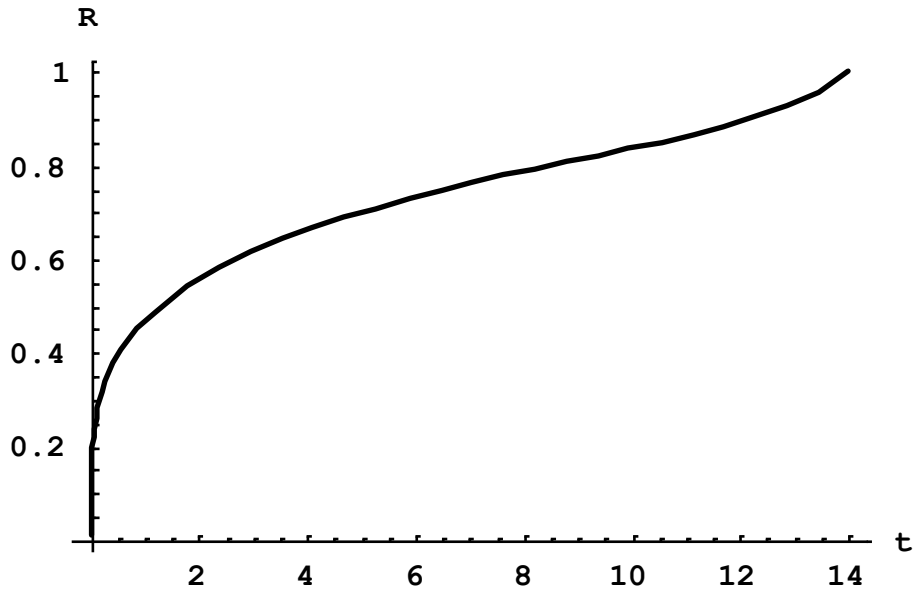

Fig. 3. Relative size of the Universe (normalized to the present size) versus the time t (in units of billion years). The inflexion point, manifesting the transition from the epoch of the decelerated expansion to the epoch of the accelerated expansion, is at about 9 billion years.

## 5. Conclusions

There is a garden variety of hypotheses that, while attempting to explain dark matter, go beyond the Standard Model. Some of them propose never-discovered subatomic particles, while others propose to significantly change the known physical laws. Out of these hypotheses, the one that came relatively close to having an experimental confirmation (by Adlarson et al (2014)), is the dibaryon hypothesis. However, Bugg (2014) pointed out logical flaws in the interpretation of that experiment as the discovery of the dibaryon and provided an alternative explanation to the experimental results of Adlarson et al (2014).

There is only one explanation of dark matter that has the following four features simultaneously:
1) it has the experimental confirmation, namely from the analysis of atomic experiments (Oks 2001);
2) it does not go beyond the Standard Model (and thus is favored by the Occam razor principle);
3) it is based on the standard quantum mechanics, namely on the Dirac equation – without any change to the physical laws (and thus is favored by the Occam razor principle);
4) it explains the perplexing astrophysical observations by Bowman et al (2018), as well as the puzzling astrophysical observation by Jeffrey et al (2021)
This is the explanation based on the existing second flavor of hydrogen atoms (Oks 2020a, 2020b, 2021b).

This explanation belongs to the category of baryonic dark matter. The results by Bowman et al (2018) and Barkana (2018), constituted a clear proof – according to the analysis by

McGaugh (2018) – that dark matter is baryonic and hypotheses on non-baryonic nature of dark matter have to be excluded.

The situation with dark energy is less certain. The least exotic and the most popular hypothesis is the fifth fundamental force – quintessence, proposed by Peebles & Ratra (1988), which is a kind of a scalar field. Within the Standard Model, scalar fields must have relatively large masses, so that viable quintessence models should go beyond the Standard Model, as noted above. This hypothesis is popular, but whether the dark energy is quintessence is not decided by the popularity contest.

Besides, there are hypotheses that dark energy does not exist. These hypotheses – such as modified gravity etc, or the one by Oks (2021b) based on the application of Dirac's Generalized Hamiltonian Dynamics – provide alternative explanations to the entire time evolution of the Universe

**Funding.** This research receive no external funding.

**Conflict of interest.** The author declares no conflict of interest.